\documentclass[fleqn,usenatbib]{mnras}

\usepackage{newtxtext,newtxmath}
\usepackage[T1]{fontenc}
\usepackage{ae,aecompl}

\usepackage{graphicx}	% Including figure files
\usepackage{amsmath}	% Advanced maths commands
\newcommand{\ms}{\mbox{m\,s$^{-1}$}}
\newcommand{\kms}{\mbox{km\,s$^{-1}$}}

\newcommand{\Mjup}{\mbox{$M_{\rm Jup}$}}

\newcommand{\Mearth}{\mbox{$M_{\oplus}$}}

\newcommand{\gtsimeq}{\raisebox{-0.6ex}{$\,\stackrel
        {\raisebox{-.2ex}{$\textstyle >$}}{\sim}\,$}}

\title[Optimising Radial Velocity Detection Limits]{Optimising Radial Velocity Detection Limits for Southern Habitable Worlds Observatory Targets}

\author[R.A. Wittenmyer et al.]{
Robert A. Wittenmyer,$^{1}$\thanks{E-mail: rob.w@unisq.edu.au (RW)}
Adriana Errico,$^{1}$ 
Timothy R. Holt,$^{1}$
Jonathan Horner,$^{1}$
\newauthor Caleb K. Harada,$^{2}$\thanks{National Science Foundation GRFP Fellow}
Stephen R. Kane,$^{3}$
Zhexing Li,$^{3}$
Tara Fetherolf$^{3}$
\\
$^{1}$Centre for Astrophysics, University of Southern Queensland, Toowoomba, QLD 4350 Australia\\
$^{2}$Department of Astronomy, 501 Campbell Hall \#3411, University of California, Berkeley, CA 94720, USA\\
$^{3}$Department of Earth and Planetary Sciences, University of California, Riverside, CA 92521, USA\\
}

\date{Accepted XXX. Received YYY; in original form ZZZ}

\pubyear{2024}

\begin{document}
\label{firstpage}
\pagerange{\pageref{firstpage}--\pageref{lastpage}}
\maketitle

\begin{abstract}

The planned NASA Habitable Worlds Observatory (HWO) flagship mission aims to image and spectroscopically characterise 25 Earth-size planets in the habitable zones of their stars.  However, one giant planet in the habitable zone can ruin your whole day.  Recent work has examined the current state of our knowledge on the presence or absence of such objects in samples of likely HWO targets, and that knowledge has been found wanting; even Saturn-mass planets remain undetectable in many of these systems.  In this work, we present simulations assessing the degree to which new campaigns of high-cadence radial velocity observations can ameliorate this woeful state of affairs.  In particular, we highlight the value of moderate-precision but highly flexibly-scheduled RV facilities in aiding this necessary HWO precursor science.  We find that for a subset of Southern HWO stars, 6 years of new RVs from the Minerva-Australis telescope array in Australia can improve the median detection sensitivity in the habitable zones of 13 likely HWO targets to $\sim$50\Mearth, an improvement of $\sim$44\%.  % updated to the crappier reality

\end{abstract}


%%%%%%%%%%%%%%%%%%%%%%%%%%%%%%%%%%%%%%%%%%%%%%%%%%

%%%%%%%%%%%%%%%%% BODY OF PAPER %%%%%%%%%%%%%%%%%%

\section{Introduction} \label{sec:intro}

For generations, humanity has yearned to understand our place in the cosmos. Are we alone? Is the Solar system unique? Could there be planets like our home orbiting distant stars?

Until the past few decades, all we could do was speculate. As recently as the early 1990s, we knew of just a single planetary system -- the Solar system, and there remained significant debate over whether planets would one day be found to be common around other stars, or if the Solar system was a peculiar anomaly, alone in a vast, uncaring cosmos.

% Alone! I'm alone! I'm a lonely, insignificant speck on a has-been planet orbited by a cold, indifferent sun!

The first hints that planets might be common in the cosmos came in the early 1980s, with the launch of the Infrared Astronomical Satellite (IRAS). That spacecraft revealed stars that were brighter than expected at infrared wavelengths \citep{Vega,BetaPic,AumDisk,IRASDisks}. That excess infrared radiation was the result of vast swathes of debris orbiting those stars -- the first detected debris disks -- and was clear evidence that the process by which planets form was in action around other stars.

In the latter years of the 20th Century, astronomers finally discovered the first planets orbiting main-sequence stars other than the Sun \citep[e.g.][]{early1,early2,early3}. To the great surprise of the astronomical community, the planetary systems so revealed bore precious little relation to our own home.

The decades since have seen an incredible growth in the number of known exoplanets -- led primarily by the great space observatories \textit{Kepler} \citep[e.g.][]{Kepler1,Kepler2,Kepler3,Kepler4} and \textit{TESS} \citep[e.g.][]{TESS1,TESS2,TESS3,TESS4}. But whilst the number and variety of planetary systems we have discovered have continued to grow rapidly, there remains a remarkable dearth of systems with architectures like our own \citep[e.g.][]{witt11,witt14,Witt16,ag18,bonomo23}\footnote{For a detailed overview of our understanding of the Solar system in the context of exoplanetary science, we direct the interested reader to \cite{SSRev}, and references therein.}. This poses an obvious question -- where are all the Solar system analogues?

The question of whether planetary systems like our own are common, scarce, or incredibly rare, is particularly important in the context of humanity's efforts to answer the question `are we alone?'. In the coming years, the search for life beyond the Solar system will begin in earnest, and significant work has been undertaken in efforts to help direct that search to the most promising targets \citep[e.g.][]{menou03,witt09,FoF1,FoF2,FoF3,HornerJones10,FoF4,Vervoort,kane24a}. 

Planned for the 2040s, the Habitable Worlds Observatory (HWO) is recommended \citep{decadal} as a NASA flagship mission to launch a 6-8m ultraviolet/optical/infrared space telescope with the primary mission of directly imaging Earth-size planets in the habitable zones of their stars.  The goal of HWO will be to obtain images and spectra of the scattered light from 25 such exo-Earths. The angular separation constraints of direct imaging impose stringent limits on the distance to candidate target stars.  Hence, the likely targets are predominantly nearby bright stars that have, in the main, been extensively studied by long-term radial-velocity exoplanet surveys.  Occurrence-rate studies have shown that approximately 10\% of Solar-type stars host giant planets beyond 1\,au \citep[e.g.][]{fernandes19, witt20, fulton21, bonomo23}.  Such objects are of course dynamically incompatible with a terrestrial planet in that region, which coincides with the habitable zone for these types of stars. 

Recent work by \citet{l23} (hereafter L23) assessed the detectability of potentially disruptive habitable-zone interlopers in a sample of likely Southern HWO target stars.  Since candidate HWO targets are preferentially nearby, bright, Solar-type stars, many are well-studied from legacy radial velocity (RV) surveys going back decades.  L23 gathered all available RV data on these stars, analysed them for signals from planets and stellar activity, then performed extensive injection-recovery tests to determine the current detection sensitivity at the ``Earth Equivalent Insolation Distance'' (EEID).  The results included this damning statement: ``for many of these stars we are not yet sensitive to even Saturn-mass planets in the habitable zone, let alone smaller planets, highlighting the need for future EPRV [extreme precision RV] vetting efforts.'' Moreover, the dynamical simulations for the HWO known exoplanet systems by \citet{kane24b}, and subsequent RV assessment by \citet{kane24c}, demonstrated the critical importance of detecting planets as small as Neptune that can serve as sources of dynamical instability in the habitable zone.

In this paper, we explore observing strategies with the aim of improving the detection sensitivities derived in L23.  In particular, we seek to quantify the degree to which small, flexibly-scheduled telescopes can contribute to this effort.  It is obvious that intensive RV campaigns observing the HWO target stars with the world's most precise RV instruments would handily resolve the distressing state of affairs revealed by L23.  But such an approach is inefficient; we seek to identify those stars which would benefit most from attention with 1-2m class ``regular RV'' facilities in an effort to inform better allocation of limited EPRV resources.  We examine the benefit to sensitivity achieved by campaigns at various levels of intensity (observing cadence).  Section 2 describes the simulated RV data properties, Section 3 gives our results and discussion, and we conclude in Section 4.

%-----------------------------------------------------------------------
\section{Simulation setup} \label{sec:style}

\begin{figure}
\includegraphics[width=\columnwidth]{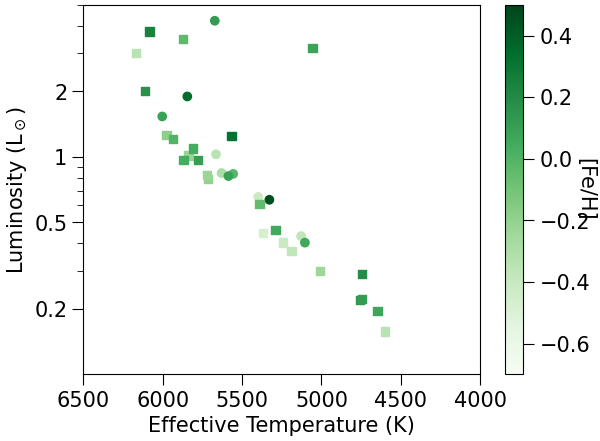}
\caption{Hertzsprung–Russell diagram for the 36 stars in our sample.  Stellar metallicity is represented by the colour bar, and the circular points indicate those stars that are currently known to harbor planets.}
\label{fig:hr}
\end{figure}

We choose 36 Southern stars from L23 that are also included in the most recent NASA Exoplanet Exploration Program Mission Star List given in \citet{mamajek24}.  Those stars are listed in Table~\ref{tab:targets} along with their Earth Equivalent Insolation Distance (EEID) and the current RV detection sensitivity, as given in Table 12 of L23. Those detection limits are given here and throughout this work as the mass for which 50\% of injected planets were successfully recovered. Shown in Figure~\ref{fig:hr} is a Hertzsprung–Russell diagram for our stellar sample, where the colour of the data points represents the stellar metallicity. Those data points shown as circles are stars presently known to harbor planets. Such visualization of the stellar sample summarizes the breadth of the stellar properties, and the metallicity may be indicative of the likelihood for additional planets being present in those systems \citep{fv05,buchhave2014,brewer2018b,buchhave2018}.

We consider three values of observing cadence $C$: 5, 10, and 20 days.  We then generate the simulated observation times as follows: Starting at an arbitrary date JD=2460000.0, the time until the next observation is drawn from a Gaussian distribution with a mean of $C$ and width $\sigma=C/3$, where $C$ is the desired cadence in days and has values of ${5, 10, 20}$.  This arrangement simulates a 1/3 weather loss by imposing the stochasticity expected from real observing conditions.  Seasonality is simulated by forbidding observations during a 4-month period each year.

This setup ensures that all the simulations share a similar temporal baseline (i.e. 3 or 6 years of new observations) as well as the number of new observations within each test case of observing cadence. However, such a setup creates a scenario where the time difference between the last real observation and the new data will vary for each star. As mentioned in Li et al. (2025, submitted), such a time difference (temporal gap) between the last observation and future data does have a significant effect on the orbital ephemerides of planets and therefore the derived RV sensitivity as well. In particular, simulations with a larger temporal gap typically exhibit better orbital constraints and may consequently produce a higher sensitivity towards the low planetary mass regime. However, the effect such a variable has on our simulation is beyond the scope of this work and we leave that to interested readers.

%the star has a probability of $1/C$ of being observed on any given date, where $C$ is the desired cadence in days and has values of ${5, 10, 15, 20}$.  This process is repeated for each simulated day up to the total duration of observations.  Seasonality is simulated by forbidding observations during a 4-month period each year.  If a star is observed on a given date, the exact observation time is shifted by a draw from a Gaussian distribution with width 4 hours to prevent exact integer-day cadence.  The result is that a desired observing frequency is reproduced with the stochasticity expected from real observing conditions.    

Next, the simulated radial velocity (RV) is derived as follows: For each of the 36 stars, we have the real RV from archival data as compiled for the sensitivity simulations of \citet{l23}.  For stars that host planets, the known planet orbits and any trends are fitted and removed, and the residual planet-free data are used for this step.  The simulated RV data point is then drawn at random (with replacement) from the real RV for each star.  In this way, we capture the intrinsic differences in the noise level between stars; not all stars are equally well-behaved.  Often the legacy RV data come from very precise instruments such as Keck/HIRES, HARPS, or Magellan/PFS.  To properly account for the fact that here we are simulating future observations with a less-precise instrument, we add scatter to the simulated RV observation.  The RV value is ``bumped'' by an amount drawn from a Gaussian distribution with zero mean and $\sigma=4$\,\ms.  The RV uncertainty on each simulated point is then drawn from a Gaussian distribution with a mean of 4\ms\ and width of 1\ms.  This is chosen as representative of the RV precision delivered by MINERVA-Australis for typical bright Solar-type stars like those in the HWO target list.  MINERVA-Australis \citep{addison19} is used here as an exemplar ``regular RV'' facility with flexible scheduling.  It is comprised of four 0.7m Planewave CDK-700 telescopes fibre feeding a single environmentally-stabilised Kiwistar R4-100 spectrograph \citep{barnes12}.  It has been wholly dedicated to RV follow up and mass measurement of candidate planets from the \textit{Transiting Exoplanet Survey Satellite} (TESS) mission since 2019, contributing data to the confirmation of nearly 40 TESS planets to date \citep[e.g.][]{min1,min2,min3,min4}.

% revise table 1 to merge in with 2
% star, rms, EEID, sensitivity L23, Time gap, 5d, 10d, 20d 

\begin{table*}
    \centering
    \begin{tabular}{lrrrrrrr}
    \hline
    Star   & RMS    & EEID & L23 Sensitivity & Time gap & 5d & 10d & 20d \\
%    \hline
      HD  & \ms     & (au) & (\Mearth) & days & \multicolumn{3}{c}{Detection limit at EEID (\Mearth)}  \\
    \hline 
693     & 3.08 &      1.731  &  403.8 & 4236 &  92.5 & 162.9 & 227.6 \\
1581    & 4.60 &      1.123 &   9.7 & 2028 & 22.1 & 24.5 & 26.6 \\
2151    & 2.75 &      1.864 &  44.8 & 2654 & 89.6 & 68.6 & 81.8  \\
4628    & 2.74 &      0.548 &  13.4 & 2034 & 16.7 & 17.2 & 19.5 \\
7570    & 6.30 &      1.415 &  88  & 2652 & 94.0 & 122.9 & 94.6  \\
14412   & 4.31 &      0.668 &  24.5 & 1296 & 35.6 & 33.6 & 46.9  \\
20766   & 10.28 &      0.891 &  81  & 2651 & 113.3 & 163.0 & 154.4 \\
20794   & 1.62 &      0.809 &  7.4 & 787 & 35.5 & 42.9 & 36.5 \\
20807   & 4.82 &      1.008 &  21.4 & 1229 & 42.9 & 49.7 & 43.1  \\
23249   & 4.22 &      1.778 &  27.6 & 1921 & 56.1 & 50.0 & 59.1 \\
26965   & 1.92 &      0.658 &  11.8  & 1078 & 18.6 & 18.4 & 18.6 \\
30495   & 14.02 &      0.983 &  393.6  & 1951 &  256.4 & 574.0 & 823.5 \\
32147   & 2.38 &      0.539 &  9.1 & 786 & 11.7 & 10.6 & 11.2  \\
38858   & 3.75 &      0.909 &  17.2  & 1983 & 40.8 & 37.8 & 40.2  \\
39091   & 4.35 &      1.238 &  51.1  & 1115 & 33.9 & 38.5 & 30.5 \\
50281   & 7.03 &      0.469 &  55.9 & 818 & 44.1 & 56.6 & 82.9  \\
69830   &  1.29 &     0.779 &  7.4 & 1887 &  10.1 & 10.0 & 8.8 \\
72673   &  3.75 &     0.635 &  9.6 & 1851 & 17.9 & 19.1 & 18.3 \\
75732   &  4.62 &     0.797 &  35  & 1971 & 38.4 & 47.2 & 37.7  \\
76151   &  9.53 &     0.985 &  17394  & 3744 & 135.5 & 219.9 & 374.0 \\
100623  &  7.05 &    0.608 &  19.2 & 1111 & 32.9 & 26.6 & 36.9 \\
102365  &  3.20 &     0.919 &  13.6  & 1473 & 15.0 & 17.0 & 13.5 \\
102870  &  5.00 &     1.941 &  201.6  & 1396 & 782.0 & 4103 & 4113 \\
114613  &  5.14 &     2.055 &  53.6  & 640 & 95.9 & 97.3 & 103.4 \\
115617  &  3.27 &    0.914 &  15.2  & 2027 & 21.8 & 18.9 & 20.6 \\
131977  &  6.24 &     0.472 &  171.2 & 5108 & 76.6 & 129.6 & 218.3 \\
136352  &  3.86 &     1.014 &  9.7  &  2031 & 15.2 & 16.9 & 18.9 \\
140901  &  11.88 &     0.904 &  95.3  & 1078 & 134.5 & 203.8 & 177.3  \\
146233  &  9.16 &     1.046 &  17.8  & 1790 & 132.1 & 162.7 & 159.6 \\
149661  &  8.51 &     0.680 &  110  & 2768 & 136.8 & 155.0 & 244.0  \\
156026  &  3.96 &     0.397 &  13.5  & 668 & 20.4 & 23.7 & 25.2  \\
160691  &  3.82 &     1.378 &  27.7  & 2743 & 40.6 & 45.9 & 46.4 \\
190248  &  3.83 &     1.118 &  10.9  & 2338 & 17.0 & 17.6 & 16.1 \\
192310  &  2.78 &     0.636 &  7.8  & 1901 & 10.8 & 9.5 & 10.2 \\
207129  &  5.63 &     1.099 &  35.8  & 1256 &  54.2 & 59.6 & 61.0 \\
216803  &  14.00 &     0.443 &  176.5 & 2438 & 106.5 & 85.9 & 133.1 \\
\hline
    \end{tabular}
    \caption{Target list with Earth Equivalent Insolation Distance (EEID) and RV sensitivity results reprised from Table 12 of \citet{l23}, the gap in time between the last real observation and the start of our simulated observations, and the resulting sensitivities after adding a further three years of new data at three different values of observing cadence. }
    \label{tab:targets}
\end{table*}

%-----------------------------------------------------------------------
\section{Results and Discussion}

With simulated data in hand -- three cadence scenarios for each target -- we determined detection sensitivities using \texttt{RVSearch} \citep{rvsearch}.  First we test the effect of adding a further three years of new observations to the existing legacy RV data that were analysed in L23.  Table~\ref{tab:targets} shows the detection sensitivity results for the three cadences.  Intuitively, one would expect a faster cadence (5 days) to deliver better sensitivities than slower (20 days).  This is generally true, but we also see some stars where the detection limit is relatively insensitive to the observing cadence.  We interpret this as a ``floor'': new data at the moderate (4\,\ms) single-measurement precision in the presence of stellar noise does little to improve the overall sensitivity to lower-mass planets.  These are stars for which the existing data are already quite good, so further improvement must be obtained by other means, e.g. with higher-precision instruments and/or a detailed treatment of stellar activity noise.  

\begin{figure}
\includegraphics[width=\columnwidth]{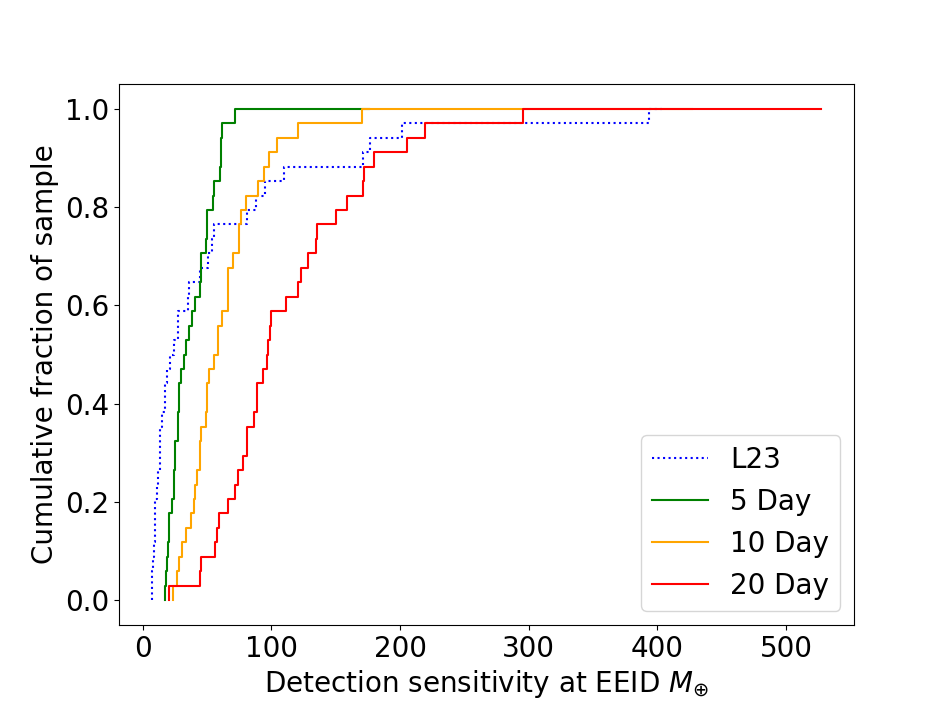}
\caption{Cumulative distribution function of all simulation results, considering 6 years of new RV data only.  Since these results only used new RV data and not the legacy data analysed in L23, the lower-cadence (20-day and 10-day) scenarios sometimes performed worse than in L23. }
\label{fig:cdf}
\end{figure}

Table~\ref{tab:targets} also features some targets where, perversely, adding 3 years of new data delivers a worse sensitivity result than that of L23.  We attribute this counterintuitive outcome to some pathologies of data sampling; in particular, the long time gap between the old and new observations can introduce strong aliases in the log-likelihood periodograms that \texttt{RVSearch} uses to identify injected periodicities.  We thus next performed the same tests with 3 and 6 years of only new simulated data, in an effort to mitigate the deleterious effects of haphazard sampling and data inhomogeneities.  Those results are given in Table~\ref{tab:3and6years}.  Here we see that the 3-year results are again sometimes worse than L23 when the EEID is at a distance such that 3 years of observations cover not much more than a single orbital period.  For those situations, we would expect the sensitivity to fall off dramatically as the orbital period of interest approaches the data span.  Hence, we hereafter consider and discuss only the 6-year simulations to obviate this artefact.  A cumulative distribution plot of the 6-year simulations is shown in Figure~\ref{fig:cdf}, comparing these results to those of L23 over the entire sample.   

\begin{figure}
\includegraphics[width=\columnwidth]{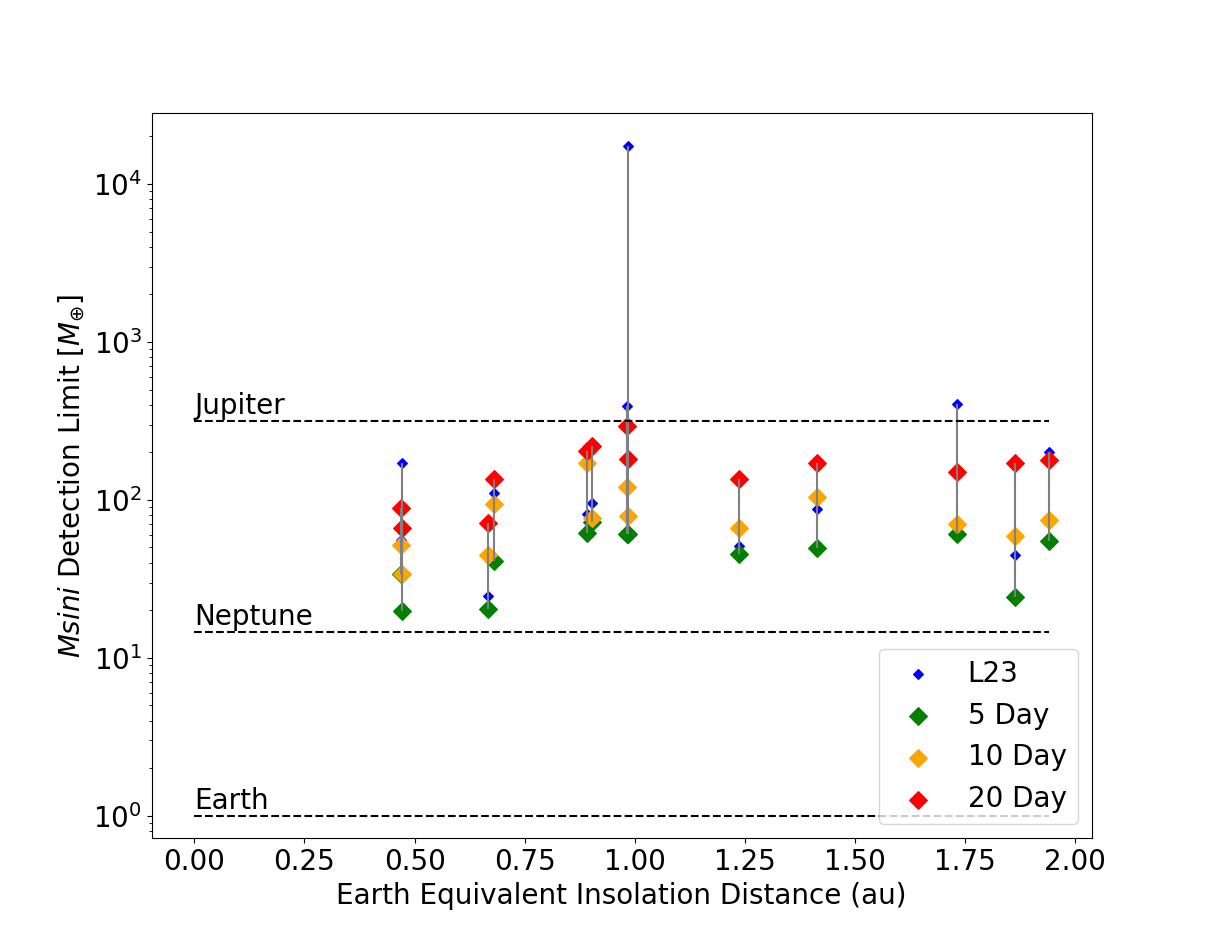}
\caption{Detailed results for the 13 stars where six years of moderate-precision monitoring significantly improves the detection sensitivity at the EEID. }
\label{fig:improvement}
\end{figure}

We next identify a subset, 13 of these 36 stars marked in bold in Table~\ref{tab:3and6years}, where the advantage of cadence outweighs the shortcoming of only moderate single-measurement precision: those for which small, flexibly-scheduled telescopes can make an improvement in detection sensitivity.  While the ultimate goal is to clear these systems of disruptive giants, to Neptune mass and below, we recognise the intrinsic limitation of the moderate-precision facilities considered herein.  As a figure of merit, a Neptune-mass planet (15\Mearth) at 1\,au imposes a radial-velocity signal of amplitude $K\sim$1.5\,\ms\ for a Solar-mass star.  Simply put, this is not possible for the Minerva-like data we have simulated here.  A Saturn-mass planet (95\Mearth) would produce an RV signal of $K\sim$9\,\ms.  Figure~\ref{fig:improvement} shows the results of our 6-year simulations compared with those of L23 for these 13 stars where this sort of higher-cadence, lower-precision data can fill a valuable niche.  The median EEID sensitivity for these stars in L23 was 95.3 \Mearth, improving to 74.7\Mearth (i.e. a 22\% improvement) after a six-year Minerva-Australis observing campaign with 10-day cadence.  Increasing to a 5-day cadence gets us 49.8\Mearth, a 48\% improvement.  For these stars, then, moderate-precision facilities can make a valuable contribution to HWO precursor science by bringing down the sensitivity limits for the least well-characterised stars on the likely target list.

%did not deliver substantial further improvement, which we attribute to a combination of intrinsic stellar activity and the noise floor of the instrument. 

% maybe can explain some of this away to "heterogeneous recovery plots" it's not always smooth and the EEID is a specific point.  What is the typical uncertainty?  no idea! RVsearch seems to repeat its 'random' draw...

% updated 2025 Jan 16. 
\begin{table*}
    \centering
    \begin{tabular}{lrrrrrrrr}
    \hline
%    Star   & EEID  & L23 result  & 5d & 10d & 20d & 5d & 10d & 20d \\
                & \multicolumn{6}{c}{Detection limit at EEID (\Mearth)}  \\
    Star       & EEID  & L23 Sensitivity  & \multicolumn{3}{c}{3 years} & \multicolumn{3}{c}{6 years} \\
    HD          & au  & \Mearth   & 5d & 10d & 20d & 5d & 10d & 20d \\
    \hline 
\textbf{693}  &    1.731 &  403.8   &  148.0  &   138.2  &   521.1  &    61.1  &    70.6  &    150.5  \\
\textbf{2151}  &  1.864 &  44.8   &   190.5  &   285.8  &   578.6  &    24.2  &    58.7  &   171.1  \\
\textbf{7570}  &  1.415  & 88.0  &  136.3  &   202.6  &   265.6  &    49.8  &   104.7  &   172.1  \\
\textbf{14412}  &   0.668  & 24.5  &   26.4  &    59.8  &   133.0  &    20.5  &    44.9  &    71.9  \\
\textbf{20766}  &  0.891  & 81.0  &  148.6  &   206.0  &   335.3  &    61.6  &   170.2  &   205.4  \\
\textbf{30495}  &  0.983  & 393.6  &  156.4  &   324.2  &   623.1  &    60.6  &   120.9  &   295.6  \\
\textbf{39091}  &  1.238  & 51.1  &   69.4  &   101.2  &   122.4  &    45.7  &   66.7   &   135.1   \\
\textbf{50281}  &  0.469  & 55.9  &   35.3  &    64.3  &   178.0  &    34.0  &    51.8  &    89.2   \\
\textbf{76151}  &  0.985  & 17394  &   83.9  &   172.0  &   393.8  &    60.6  &    78.6  &   180.5   \\
\textbf{102870}  &  1.941  &  201.6  &  249.7  &   258.8  &   894.9  &    55.0  &    74.7  &   180.2   \\
\textbf{131977}  &   0.472  & 171.2  &  42.9  &   109.8  &    96.0  &    19.7  &    34.1  &    66.4   \\
\textbf{140901}  &   0.904  & 95.3  &  83.4  &   198.4  &   446.6  &    72.0  &    76.5  &   219.1  \\
\textbf{149661}  &   0.680  & 110.0  &  66.9  &   118.4  &   222.0  &    41.1  &    94.5  &   135.8  \\
\hline
  1581  &  1.123  & 9.7  &   49.7  &   150.5  &   176.7  &    27.7  &    55.2  &   128.2  \\
  4628  &  0.548  & 13.4  &   40.9  &    67.8  &   114.5  &    18.0  &    27.1  &    59.5  \\
 20794  &  0.809  & 7.4  &   26.8  &    84.2  &   173.7  &    18.8  &    50.0  &    97.7  \\
 20807  &  1.008  & 21.4  &   40.6  &   127.9  &   330.1  &    27.4  &    66.6  &    99.6  \\
 23249  &  1.778  & 27.6  &  157.2  &   470.7  &   280.0  &    50.1  &    98.4  &    93.6  \\ 
 26965  &  0.658  & 11.8  &   44.7  &    74.6  &   148.0  &    22.9  &    30.4  &    44.9  \\
 32147  &  0.539  & 9.1  &   20.2  &    60.8  &    86.8  &    25.0  &    24.0  &    45.4  \\
 38858  &  0.909  & 17.2  &   54.5  &    84.0  &   192.5  &    28.3  &    74.6  &    86.4  \\
 69830  &  0.779  & 7.4  &   37.8  &    58.4  &    85.8  &    28.1  &    40.8  &    89.1   \\
 72673  &  0.635  & 9.6  &   38.2  &    35.3  &   120.6  &    20.8  &    40.1  &    78.4  \\
 75732  &  0.797  & 35.0  &   36.1  &    78.7  &   144.1  &    45.2  &    50.0  &    74.3   \\
100623  &  0.608  & 19.2  &   65.3  &    68.4  &   196.2  &    44.3  &    61.8  &    96.7  \\
102365  &  0.919  & 13.6  &   53.9  &   78.7  &   236.9  &     25.6  &     42.2  &    81.0   \\
114613  &  2.055  & 53.6  &  479.5  &    77.8  &   692.8  &   176.1  &   426.8  &   526.8  \\ % dafuq?
115617  &  0.914  & 15.2  &   40.5  &    58.9  &   160.0  &    32.1  &    44.8  &    81.0   \\
136352  &  1.014  & 9.7  &   49.9  &    72.2  &   121.9  &    29.6  &    49.4  &    20.3   \\ % wtaf
146233  &  1.046  & 17.8  &   40.9  &    69.5  &   201.9  &    38.7  &    45.4  &   120.7  \\
156026  &  0.397  & 13.5  &   20.2  &    31.0  &    79.9  &    17.6  &    28.7  &    56.2  \\
160691  &  1.378  & 27.7  &   33.5  &   101.6  &   229.3  &    49.4  &    58.8  &    111.5   \\
190248  &  1.118  & 10.9  &   51.8  &   130.1  &   209.0  &    35.9  &    80.7  &    99.0  \\
192310  &  0.636  & 7.8  &   38.1  &    66.1  &   116.7  &    24.6  &    37.8  &    58.2   \\
207129  &  1.099  & 35.8  &   62.3  &   131.5  &   232.2  &    55.2  &    66.6  &   122.9  \\
216803  &  0.443  & 176.5  &   86.2  &    81.0  &   265.1  &    60.3  &    89.5  &   158.5   \\
\hline
    \end{tabular}
    \caption{RV detection limit at the EEID considering \textit{only} 3 and 6 years of new data at three different values of observing cadence.  The 13 stars in boldface are those for which this strategy improves on the results of L23. }
    \label{tab:3and6years}
\end{table*}

% updated with new values 16 March.

%-----------------------------------------------------------------------
\section{Conclusions}

In this work, we have shown that small, flexibly-scheduled, moderate-precision RV facilities can make important contributions to the necessary precursor science for HWO target optimisation.  These facilities include MINERVA-Australis (as detailed in this work), MINERVA-North \citep{swift15, wilson19}, and the 1m Stellar Observations Network Group telescopes \citep{2007CoAst.150..300G, frank2017}.  We have also demonstrated that for some stars, significant improvement in the detection sensitivity is possible with e.g. a 5 to 10 day observing cadence; this is eminently feasible for these facilities.  We also note that such observations on timescales of order 3-6 years would detect, ``for free,'' RV trends attributed to giant planets moving on orbits $a\gtsimeq$3\,au.  Such cold giants may not in and of themselves disqualify potential HWO targets (except those on problematically eccentric orbits), but their presence (or absence) is a key data point in terms of understanding the overall system architecture, dynamical history, and volatile delivery regime experienced by any inner planets \citep[e.g.][]{SSRev, childs22, ogihara23, kane24a}.

Recent work by \citet{harada24b} performed a similar analysis as L23, focusing on 90 potential HWO target stars including Northern targets \citep{mamajek24, harada24a} observed with HIRES and/or HARPS.  They found a median sensitivity of $M_{\rm p} \sin i \simeq 66 \,{\rm M_\oplus}$ in the middle habitable zones, with a similarly large dispersion as L23.  They also found that the legacy HARPS and HIRES data were biased toward cooler GKM stars, motivating the continued importance of moderate-precision RV facilities in studying hotter stars that are less amenable to EPRV observations.  That work again highlights the need for further concentrated observational efforts for some heretofore neglected stars.

This work and that of L23 and \citet{harada24b} all point to the necessity to better understand the cohort of nearby Solar-type stars that will be the best targets for future imaging missions.  Given that the occurrence rate of giant planets ($M>0.3$\Mjup) steeply increases near 1\,au, from $\sim$1\% to $\sim$10\% \citep{witt20}, it is wise to characterise the potential entourage of planets that may be accompanying these stars.  The best time to start was 20 years ago; the second best time is now.

%-----------------------------------------------------------------------
\section{Data Availability} 

The data underlying this article are available on request to the corresponding author.

%-----------------------------------------------------------------------
\section*{Acknowledgements}

\textsc{Minerva}-Australis is supported by Australian Research Council LIEF Grant LE160100001, Discovery Grants DP180100972, DP220100365, and DP250101273, Mount Cuba Astronomical Foundation, and institutional partners University of Southern Queensland, UNSW Sydney, MIT, Nanjing University, George Mason University, University of Louisville, University of California Riverside, University of Florida, and The University of Texas at Austin. We acknowledge support from the NASA Astrophysics Decadal Survey Precursor Science (ADSPS) program under Grant Number 80NSSC23K1476. C.K.H.\ acknowledges support from the National Science Foundation (NSF) Graduate Research Fellowship Program (GRFP) under Grant No.~DGE 2146752.

We respectfully acknowledge the traditional custodians of all lands throughout Australia, and recognise their continued cultural and spiritual connection to the land, waterways, cosmos, and community. We pay our deepest respects to all Elders, ancestors and descendants of the Giabal, Jarowair, and Kambuwal nations, upon whose lands the \textsc{Minerva}-Australis facility at Mt Kent is situated.

%%%%%%%%%%%%%%%%%%%%%%%%%%%%%%%%%%%%%%%%%%%%%%%%%%



%\bibliographystyle{mnras}

% I'll be dead in the cold cold ground before I recognise bibtex. get off my lawn

%%%%%%%%%%%%%%%%%%%%%%%%%%%%%%%%%%%%%%%%%%%%%%%%%%

%%%%%%%%%%%%%%%%% APPENDICES %%%%%%%%%%%%%%%%%%%%%

%\appendix


%If you want to present additional material which would interrupt the flow of the main paper, it can be placed in an Appendix which appears after the list of references.

%%%%%%%%%%%%%%%%%%%%%%%%%%%%%%%%%%%%%%%%%%%%%%%%%%

% Don't change these lines
\bsp	% typesetting comment
\label{lastpage}
\end{document}